\documentclass[11pt]{article}

\usepackage[usenames,dvipsnames,svgnames,table]{xcolor} 
\usepackage[obeyspaces,hyphens,spaces]{url}
\usepackage{jcapmod}

\usepackage{booktabs}
\usepackage[english]{babel}
\usepackage{amsmath, amssymb, amsbsy, amstext, amsthm, simplewick}
\usepackage{hyperref}
\usepackage{graphicx}
\usepackage{amsfonts}
\usepackage{enumitem}
\usepackage{caption}
\usepackage{upgreek}
\usepackage{framed}
\usepackage{tensor}
\usepackage{pifont}
\usepackage{latexsym, mathrsfs}
\usepackage{array}
\usepackage{hyperref}
\usepackage{xspace}
\usepackage{longtable}
\usepackage{multirow}
\usepackage{cases}
\usepackage{empheq}
\usepackage{bm}
\usepackage{bbm}
\usepackage{braket}
\usepackage[utf8]{inputenc}

\usepackage[load-configurations=astronomy, range-units=brackets, range-phrase=-, per-mode=reciprocal, mode=math]{siunitx}
\usepackage{threeparttable}
\usepackage{subfig}

\usepackage{ragged2e}

\usepackage{hhline}
\usepackage{array}
\newcolumntype{P}[1]{>{\centering\arraybackslash}p{#1}}
\usepackage{booktabs}
\usepackage{tikz}
\usetikzlibrary{calc,shadings,patterns,tikzmark,fadings}

\definecolor{Blue}{rgb}{0.25, 0.41, 0.88}
\definecolor{Red}{rgb}{0.92,0.,0.}
\definecolor{darkorange}{rgb}{1.0,0.549,0.}
\definecolor{cobalt}{RGB}{44, 98, 120}
\definecolor{Mathematica1}{rgb}{0.368417, 0.506779, 0.709798}
\definecolor{Mathematica2}{rgb}{0.880722, 0.611041, 0.142051}
\definecolor{Mathematica3}{rgb}{0.560181, 0.691569, 0.194885}
\definecolor{Mathematica4}{rgb}{0.922526, 0.385626, 0.209179}
\definecolor{Mathematica5}{rgb}{0.528488, 0.470624, 0.701351}
\definecolor{Mathematica6}{rgb}{0.772079, 0.431554, 0.102387}
\definecolor{Mathematica7}{rgb}{0.363898, 0.618501, 0.782349}
\definecolor{Mathematica8}{rgb}{1, 0.75, 0}
\definecolor{Mathematica9}{rgb}{0.647624, 0.37816, 0.614037}
\definecolor{plotBlue}{RGB}{94, 130, 181}
\definecolor{plotRed}{RGB}{233, 85, 54}
\definecolor{plotGreen}{RGB}{142, 176, 50}
\definecolor{plotPurple}{RGB}{135, 120, 178}

\definecolor{cornellRed}{HTML}{B31B1B}
\definecolor{cornellBlue}{HTML}{0068AC}
\definecolor{cornellGreen}{HTML}{6EB43F}

\newcolumntype{C}[1]{>{\centering\let\newline\\\arraybackslash\hspace{0pt}}m{#1}}

\def\d{{\rm d}}

\makeatletter
\newlength{\apb@width}
\newcommand{\autoparbox}[2][c]{\settowidth{\apb@width}{#2}\parbox[#1]{\apb@width}{#2}}

\makeatother

\numberwithin{equation}{section}

\def\beq{\begin{equation}}
\def\eeq{\end{equation}}

\def\bea{\begin{eqnarray}}
\def\eea{\end{eqnarray}}

\def\d{{\rm d}}

\def\beq{\begin{equation}}
\def\eeq{\end{equation}}
\def\bea{\begin{eqnarray}}
\def\eea{\end{eqnarray}}

\def\d{{\rm d}}

\def\d{{\rm d}}

\DeclareRobustCommand{\SkipTocEntry}[4]{}

\usepackage{colortbl}
\definecolor{blue2}{cmyk}{1, 0.1, 0.1, 0}

\definecolor{pyBlue}{RGB}{31, 119, 180}
\definecolor{pyRed}{RGB}{214, 39, 40}
\definecolor{pyGreen}{RGB}{44, 160, 44}
\definecolor{pyBlue2}{RGB}{0, 111, 237}
\definecolor{pyRed2}{RGB}{224, 52, 36}

\makeatletter
\def\Ddots{\mathinner{\mkern1mu\raise\p@
\vbox{\kern7\p@\hbox{.}}\mkern2mu
\raise4\p@\hbox{.}\mkern2mu\raise7\p@\hbox{.}\mkern1mu}}
\makeatother

\def\tmp{bbh}
\def\gen{g}

\begin{document}

\pagenumbering{roman}
\begin{titlepage}
\baselineskip=15.5pt \thispagestyle{empty}
\begin{flushright}

\end{flushright}
\vspace{1cm}

\begin{center}
\fontsize{19}{24}\selectfont  \bfseries Searching for General Binary Inspirals \vskip 0pt with Gravitational Waves 
\end{center}

\vspace{0.1cm}
\begin{center}
\fontsize{12}{18}\selectfont Horng Sheng Chia$^{1}$ and Thomas D. P. Edwards$^{1, 2, 3}$
\end{center}

\begin{center}
\vskip8pt

\textsl{$^1$ Institute for Theoretical Physics, University of Amsterdam,\\Science Park 904, Amsterdam, 1098 XH, The Netherlands}

\vskip8pt

\textsl{$^2$ Gravitation Astroparticle Physics Amsterdam (GRAPPA), University of Amsterdam, \\ Science Park 904, Amsterdam, 1098 XH, The Netherlands}

 \vskip8pt
 
\textsl{$^3$ The Oskar Klein Centre, Department of Physics, \\ Stockholm University,  AlbaNova, SE-10691 Stockholm, Sweden}

\end{center}

\vspace{1.2cm}
\hrule \vspace{0.3cm}
\noindent {\bf Abstract} 

\noindent We study whether binary black hole template banks can be used to search for the gravitational waves emitted by general binary coalescences.
To recover binary signals from noisy data, matched-filtering techniques are typically required. 
This is especially true for low-mass systems, with total mass $M \lesssim 10 \, M_\odot$, which can inspiral in the LIGO and Virgo frequency bands for thousands of cycles.
In this paper, we focus on the detectability of low-mass binary systems whose individual components can have large spin-induced quadrupole moments and small compactness.
The quadrupole contributes to the phase evolution of the waveform whereas the compactness affects the merger frequency of the binary.
We find that binary black hole templates (with dimensionless quadrupole $\kappa=1$) cannot be reliably used to search for objects with large quadrupoles ($\kappa\gtrsim 20$) over a wide range of parameter space.
This is especially true if the general object is highly spinning and has a larger mass than its binary companion.
A binary that consists of objects with small compactness could merge in the LIGO and Virgo frequency bands, thereby reducing its accumulated signal-to-noise ratio during the inspiraling regime.
Template banks which include these more general waveforms must therefore be constructed. These extended banks would allow us to realistically search for the existence of new astrophysical and beyond the Standard Model compact objects.



\vskip10pt

 

\hrule
\vskip10pt

\end{titlepage}

\thispagestyle{empty}
\setcounter{page}{2}
\tableofcontents

\newpage
\pagenumbering{arabic}
\setcounter{page}{1}

\clearpage
\section{Introduction}
 \label{sec:introduction}

The direct detection of gravitational waves~\cite{Abbott:2016blz, TheLIGOScientific:2017qsa} has opened up a unique way to view the dark side of our Universe. By virtue of Einstein's equivalence principle, all forms of matter and energy density must interact gravitationally, making gravitational waves universal probes of new physics in regimes which are typically inaccessible by other experimental means. This new observational window has come at a time when challenges in fundamental physics, cosmology, and astrophysics remain abound, for instance: we still do not know what $95\%$ of the energy budget of our universe is~\cite{Aghanim:2018eyx}; there is a significant discrepancy between different cosmological measurements of the Hubble constant \cite{Verde:2019ivm}; and the origin of supermassive black holes in the early universe is still unknown \cite{Haiman:2012ic}.
The current network of gravitational-wave detectors allows us to explore large volumes of our dark universe, hopefully helping to answer some of these questions~\cite{Barack:2018yly, Bertone:2019irm}. Next generation detectors will see further and over a greater range of frequencies, revealing even more --- the future of gravitational-wave astrophysics is bright~\cite{Maggiore:2019uih, Barausse:2020rsu}.

\vskip 4pt
Compact binary systems are one of the loudest and most important sources of gravitational waves. These systems are unique in that accurate computations of their gravitational waveforms, especially in the early inspiraling regime, are attainable \cite{Blanchet:2006zz, Porto:2016pyg, Schafer:2018kuf, Bern:2019crd, Levi:2018nxp}. This makes an observed waveform a rich source of information about the binary's dynamics and the physics of its components. In fact, accurate waveform models are essential for detecting these signals, as they are most reliably extracted from noisy data streams through \textit{matching} with template waveforms~\cite{Thorne1980Lectures, Sathyaprakash:1991mt, Dhurandhar:1992mw, Cutler:1992tc}. Our reliance on this matched-filtering technique, however, also implies that we are bound to only detect signals that we can predict.\footnote{Coherent burst search methods~\cite{Klimenko:2015ypf, Lynch:2015yin, Cornish:2014kda} have been developed to detect transient gravitational waves in a model-independent way. Nevertheless, they only capture loud and short-duration events, such as the near-merger regime of binary coalescences, and are insensitive to weak and long-duration binary inspirals~\cite{Abbott:2016ezn,Abbott:2019heg, LIGOScientific:2018mvr}.} In particular, an order-one mismatch between the phases of the signal and template waveforms can easily degrade the detectability of a signal~\cite{Cutler:1992tc}, thereby resulting in a missed event. It is hence crucial that we develop increasingly precise template banks that also cover a wider range of the parameter space. This strategy would certainly broaden existing searches for binary black hole and neutron star systems. At the same time it could potentially discover new types of compact objects that have been proposed in various beyond the Standard Model (BSM) scenarios, such as: primordial black holes~\cite{HawkingPBH, Sasaki:2018dmp}, gravitational atoms~\cite{Arvanitaki:2009fg, Arvanitaki:2010sy}, boson stars~\cite{Kaup1968, Ruffini1969, Breit:1983nr, Colpi1986, Liebling:2012fv},  soliton stars~\cite{Lee1987-2, Lee1987-3, Lee1987-4, Lynn:1988rb}, oscillons~\cite{Seidel1991, Copeland:1995fq}, and dark matter spikes~\cite{Gondolo:1999ef, Ferrer:2017xwm}.

\begin{figure}[t]
        \centering
        \includegraphics[scale=1, trim=15 0 0 0]{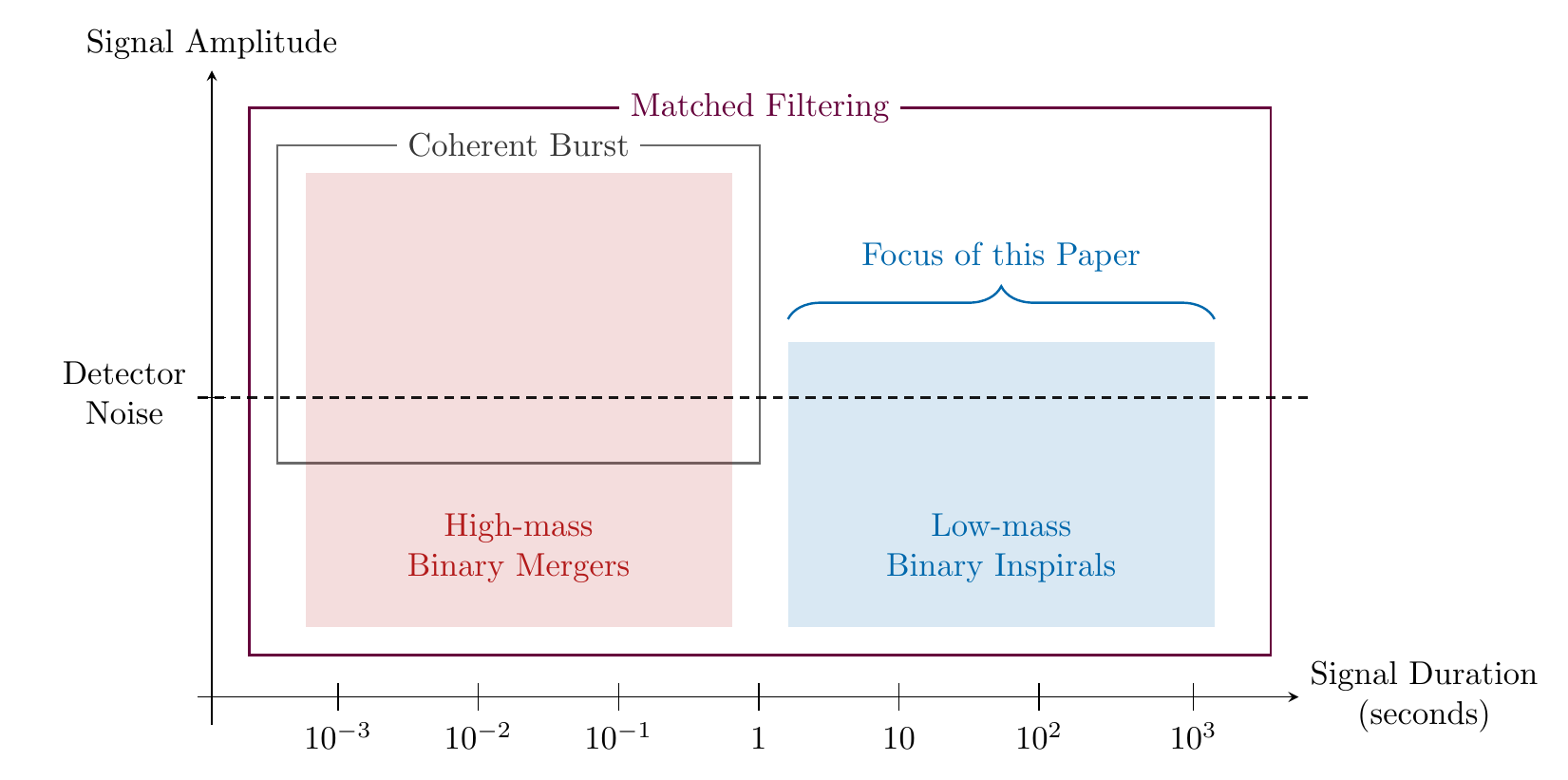}
\caption{Overview of the different methods used to search for gravitational waves emitted by a general binary system with the LIGO and Virgo detectors. The signals of low-mass binary systems are typically weaker and last longer than those of high-mass systems. The horizontal dashed line schematically illustrates the detector noise level, below which coherent burst searches become quickly insensitive. In this paper we focus on low-mass binary inspirals which are \textit{only} detectable with matched filtering. } 
    \label{fig:search_overview}
\end{figure}

\vskip 4pt

Modeling accurate waveforms for general\footnote{
In this paper, we use the word `general' to refer to any binary system involving at least one non-standard astrophysical compact object. In contrast, we use the word `standard' to refer to binary systems involving only black holes and neutron stars.} binary inspirals is a laborious task. Fortunately, the complicated microphysics of a general object are integrated over astrophysical length scales, and affects their binary waveforms through various universal \textit{finite-size effects}. For instance, different microphysics often result in distinctive mass-radius relations for the object. The radius of the object, or equivalently its compactness, in turn dictates the merging frequency of the binary~\cite{Giudice:2016zpa}. In addition, these finite-size effects induce subtle yet important phase imprints in the binary's waveform. For a non-spinning object, the dominant effect arises from the object's induced tidal deformation, which first enters the waveform at five post-Newtonian (PN) order~\cite{nrgr, Flanagan:2007ix}. While this tidal effect offers a useful way of testing the nature of the binary components~\cite{Flanagan:2007ix, Porto:2016zng, Cardoso:2017cfl, Sennett:2017etc}, it would only become appreciable near the merger of the binary, where strong gravity dynamics must be taken into account. This modeling challenge can be circumvented when considering spinning objects, which generate a series of spin-induced multipole moments~\cite{Geroch:1970cd, Hansen:1974zz, Thorne:1980ru} that perturb the dynamics of the binary in its early inspiraling regime. Specifically, the dominant spin-induced quadrupole moment first enters the phase evolution of the waveform at 2PN order~\cite{Poisson:1997ha}, and has been incorporated into existing templates for binary black hole systems. This quadrupolar term is especially significant for the BSM objects described above, as it can be orders-of-magnitude larger than those of black holes and neutron stars~\cite{Ryan:1996nk, Herdeiro:2014goa, Baumann:2018vus}. In certain examples, the time dependence of the quadrupole can provide further unique fingerprints of the masses and intrinsic spins of the boson fields that constitute the object~\cite{Baumann:2019ztm, Baumann:2019eav}. Since analytic predictions of the waveform in the early inspiral regime are known in detail, this spin-induced finite-size effect is a much cleaner probe of new physics than the tidal deformability.

\vskip 4pt

In this paper, we attempt to address the following question: to what extent can existing template banks be used to \textit{search} for gravitational-wave signals emitted by general binary coalescences? We do so by computing the effectualness~\cite{Damour:1997ub, Buonanno:2009zt} of existing template banks to general waveforms. The effectualness describes how much signal-to-noise ratio is retained when computing the overlap between a signal and the best-fitting template waveform in a bank. In addition to the usual mass and spin parameters, these general waveforms incorporate the effects of the spin-induced quadrupole moment and the compactness of the binary components. A detection of these general signals would therefore represent a discovery of new physics in binary systems. Our template bank is designed to resemble those used by the LIGO/Virgo collaboration \cite{Brown:2012qf,DalCanton:2017ala,Roy:2017oul,2018arXiv181205121M}, demonstrating that, if these new signals exist, they could remain undetected. When constructing our template bank, we follow closely the geometric-placement method presented in Refs.~\cite{Roulet:2019hzy}. Our work is complementary to Ref.~\cite{Venumadhav:2019tad, Venumadhav:2019lyq} where they further optimized the LIGO/Virgo search pipelines. Instead, we hope to broaden the searches beyond these standard binary black hole and binary neutron star signatures. 

\vskip 4pt

We emphasize that our work is in contrast to several proposed tests of new physics in binary systems~\cite{Krishnendu:2017shb, Krishnendu:2018nqa, Kastha:2018bcr, Kastha:2019brk}, which seek to measure or constrain plausible parametric deviations in observed waveforms. This \textit{a priori} assumes a successful detection of the new binary system. Detection is typically achieved through matched filtering with current template banks, which necessarily means that the waveform deviations are small. Our focus is instead on the detectability of these plausible new binary signals, including those that incur large deviations from the binary black hole template waveforms.

\vskip 4pt

The ability to generate accurate template waveforms is a crucial prerequisite to achieving our goal. We therefore restrict ourselves to general low-mass binary systems, where the total mass of the binary is $M \lesssim 10 \, M_\odot$, for the following reasons:
\begin{itemize}

    \item In the LIGO and Virgo detectors, the inspiral regime only dominates for low-mass binary systems. The binary inspiral is an interesting regime because analytic results of the PN dynamics are readily available. This provides us with a well-defined framework to construct precise waveforms that incorporate additional physics, such as finite-size effects.

    \item The inspiraling signals of these systems last up to several minutes (corresponding to hundreds or thousands of cycles) and are typically very weak. They are therefore hard to detect with coherent burst searches. Matched filtering is the optimal and only realistic avenue to search for them  (cf. Fig.~\ref{fig:search_overview} for a comparison of these different search techniques).
    
\end{itemize}
\noindent By assuming that the PN dynamics are valid up until the merger regime, we have implicitly ignored other plausible effects that may occur even in the early inspiraling regime such as: Roche-lobe mass transfer~\cite{Paczynski1971}; third-body perturbation~\cite{Kozai, LIDOV1962719, Lim:2020cvm}; floating, sinking, or kicked orbits~\cite{Baumann:2019ztm}; dark matter environmental effects~\cite{Blas:2016ddr, Edwards:2019tzf, Kavanagh:2020cfn}; new fifth forces~\cite{Hook:2017psm, Kopp:2018jom, Wong:2019yoc}; and strong gravitational dynamics~\cite{Bezares:2017mzk, Palenzuela:2017kcg, Bezares:2018qwa}. Despite these limitations, our general waveforms still capture a wide class of new types of binary systems which have been overlooked in the literature. Crucially, we believe that this work represents a fundamental step towards realistically searching for new physics in the nascent field of gravitational-wave astronomy.

\vskip 8pt

\paragraph{Outline}  The structure of this paper is as follows: in Section~\ref{sec:binarywaveform}, we construct the waveform for a general binary inspiral. Specifically, the impact of various finite-size effects on the waveform will be incorporated. In Section~\ref{sec:testbank}, we construct a template bank that is representative of those used in standard search pipelines. We then compute the effectualness of our general waveform to this template bank. Finally, we summarize and present an outlook in Section~\ref{sec:conclusion}.

\paragraph{Convention}  We work in geometric units, $G=c=1$.

\pagebreak

\section{General Inspiral Waveform} \label{sec:binarywaveform}

In this section, we construct the template waveforms for general binary inspirals. We first review how  an astrophysical object's multipole structure imprints itself on the phase of the gravitational waves emitted by the binary system (\S\ref{sec:phase}). We then describe how an object's size, parameterized through its compactness, affects the cutoff frequency of our template waveforms (\S\ref{sec:peakfreq}). These waveforms are generalizations of those constructed for binary systems with black holes and neutron stars. Their effectualness to existing template banks will be studied in Section~\ref{sec:testbank}.

\subsection{Dephasing from Quadrupole Moment} \label{sec:phase}

The shape of a general astrophysical object, when viewed at large distances, can be described through a series of source multipole moments~\cite{Geroch:1970cd, Hansen:1974zz, Thorne:1980ru}. In this paper, we only consider objects that are spherically symmetric when they are not spinning --- in this limit only the monopole contributes.\footnote{This need not be the case, as a general astrophysical object can inherit higher-order permanent multipole moments, which are present even when the object is not spinning} Birkhoff's theorem~\cite{Birkhoff, Jebsen} then implies that, regardless of the underlying microphysical properties of the object, its exterior metric is given by the Schwarzschild solution and is therefore only described by its mass, $m$. In this case, it is challenging to distinguish between different types of non-spinning objects (though we will shortly discuss how induced tidal effects can help to alleviate this degeneracy). 

\vskip 4pt

Fortunately, the nature of an astrophysical object can be readily probed when it has non-vanishing spin. In particular, its spinning motion generates a hierarchy of axisymmetric multipole moments, whose precise values \textit{do} depend on the object's microscopic properties. The dominant moment is given by the axisymmetric quadrupole, $Q$, which is often parameterized through~\cite{Poisson:1997ha}
\beq
    Q \equiv - \kappa \, m^3 \chi^2  \, , \label{eqn:kappaDef}
\eeq
 where $\chi \equiv S / m^2$ is the dimensionless spin parameter, with $S$ being the spin angular momentum, and $\kappa$ is the dimensionless quadrupole parameter, which quantifies the amount of shape deformation due to the spinning motion. In particular, the larger the (positive) value of $\kappa$, the more oblate the object is. Crucially, the value of $\kappa$ depends sensitively on the detailed properties of the object. For instance, it is known that $\kappa = 1$ for Kerr black holes~\cite{Hansen:1974zz, Thorne:1980ru}, while $2\lesssim \kappa \lesssim 10$ for neutron stars, with the precise value depending on the nuclear equation of state~\cite{Laarakkers:1997hb, Pappas:2012ns}. For more speculative objects, such as superradiant boson clouds and boson stars, $\kappa$ can be as large as $\sim 10^2 -10^3$. It can even develop oscillatory time-dependences or vary significantly with $\chi$~\cite{Baumann:2018vus, Baumann:2019ztm, Ryan:1996nk, Herdeiro:2014goa}. Absent a specific compact-object model in mind, we will henceforth treat $\kappa$ as a free constant parameter, with the requirement that $\kappa \geq 1$.

\vskip 4pt

\begin{figure}[t!]
        \centering
        \includegraphics[width=1.0\textwidth]{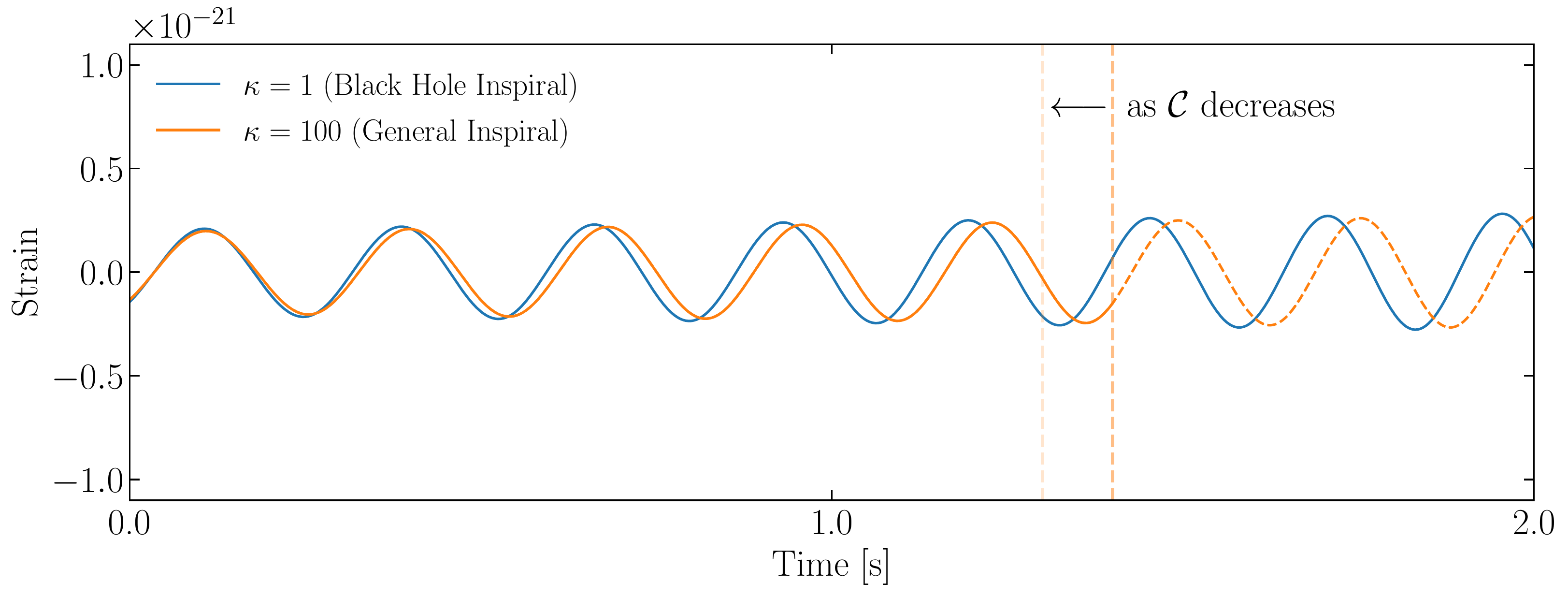}
\caption{Comparison between the waveforms of a binary black hole inspiral and a general binary inspiral. In both cases, the binary components' masses and spins are $m_1 = m_2 = 2 M_\odot$ and $\chi_1 = \chi_2 = 0.8$. No amplitude modulation is observed as we assume that the components' spins are aligned with the orbital angular momentum of the binary. One of the binary components is assumed to be a black hole $(\kappa_1 = 1)$, while the other can be a more general object $(\kappa_2 \equiv \kappa \geq 1)$. The vertical dashed line represents the cutoff of our general waveform, which happens earlier in the inspiral (lower frequency) for binary components which are less compact (i.e. as $\mathcal{C}$ decreases).}
    \label{fig:waveform}
\end{figure}

When the object is part of a binary system, the precise effect of $Q$ on the gravitational-wave signal is known in the early-inspiral, post-Newtonian regime of the coalescence~\cite{Barker75, Poisson:1997ha, Porto:2005ac, Porto:2008jj, Hergt:2010pa, Buonanno:2012rv, Bohe:2015ana, Marsat:2014xea, Krishnendu:2017shb, Levi:2015msa}. To simplify our analysis, we restrict ourselves to binary orbits which are quasi-circular, and assume that the binary components' spins are parallel to the (Newtonian) orbital angular momentum vector of the binary. In the Fourier domain, the gravitational wave strain is~\cite{Sathyaprakash:1991mt, Cutler:1994ys} 
\beq
\tilde{h} (f; \bm{p}) = \mathcal{A}(\bm{p}) \, f^{-7/6} \, e^{i \psi (f; \bm{p})} \, \theta \big( f_{\rm cut} (\bm{p}) - f \big)\, , \label{eqn:waveform}
\eeq
where $f$ is the gravitational wave frequency, $\bm{p}$ is the set of intrinsic parameters of the binary, $\mathcal{A}$ is the strain amplitude,\footnote{The amplitude $\mathcal{A}$, as defined in (\ref{eqn:waveform}), is independent of $f$ at leading Newtonian order. We will ignore higher-order PN corrections to $\mathcal{A}$, as they do not substantially affect the overlap between different waveforms. As a result, the constant $\mathcal{A}$ disappears in the normalized inner product (see Section~\ref{sec:testbank} later).} $\psi$ is the phase, $\theta$ is the Heaviside theta function, and $f_{\rm cut}$ is the cutoff frequency of our general waveform. Schematically, the phase evolution reads
\beq
\begin{aligned}
\psi (f; \bm{p})  = & \,\, 2 \pi f t_c - \phi_c - \frac{\pi}{4} + \frac{3}{128 \hskip 1pt \nu \hskip 1pt v^5} \Big( \psi_{\rm NS} + \psi_{\rm S} \Big) \, , \label{eqn:TaylorF2phase}
\end{aligned}
\eeq 
where $t_c$ is the time of coalescence, $\phi_c$ is the phase of coalescence,\footnote{More precisely, $t_c$ and $\phi_c$ are the time and phase at the cutoff frequency $f_{\rm cut}$.} $\nu = m_1 m_2 / M^2$ is the symmetric mass ratio, $M=m_1 + m_2$ is the total mass of the binary, and $v = (\pi M f)^{1/3}$ is the PN expansion parameter. The quantities $\psi_{\rm NS}$ and $\psi_{\rm S}$ represent the non-spinning and spinning phase contribution, respectively. Because the overlap between waveforms are especially sensitive to phase coherence~\cite{Cutler:1992tc}, we will retain terms in the phase up to 3.5PN order --- these terms are fully known in the literature; see for example Refs.~\cite{Arun:2004hn, Wade:2013hoa, Mishra:2016whh, Krishnendu:2017shb}. The quadrupole parameter $\kappa$ in (\ref{eqn:kappaDef}) contributes to $\psi_{\rm S}$ through the interaction between $Q$ and the tidal field sourced by the binary companion. It first appears at 2PN order~\cite{Poisson:1997ha, Wade:2013hoa}
\beq
\psi_{\rm S} \supset - 50 \sum_{i=1}^2  \, \left( \frac{m_i}{M} \right)^2 \kappa_i \hskip 1pt  \chi_i^2 \, v^4 \, , \label{eqn:Kappa2PN}
\eeq
where the subscript $i=1, 2$ represents each of the binary components. It also appears in the phase at 3PN order, though in a slightly complicated manner~\cite{Krishnendu:2017shb}
\beq
\psi_{\rm S} \supset \frac{5}{84} \sum_{i=1}^2 \sum_{j \neq i}    \, \left[ 15609  \left( \frac{m_i}{M}\right)^2 + 27032 \hskip 1pt \frac{m_i m_j}{M^2} + 9407 \hskip 1pt \left( \frac{m_j}{M} \right)^2 \right] \left( \frac{m_i}{M} \right)^2  \kappa_i \hskip 1pt \chi_i^2 \, v^6 \, . \label{eqn:Kappa3PN}
\eeq
We have simplified (\ref{eqn:Kappa2PN}) and (\ref{eqn:Kappa3PN}) by enforcing $\chi_i$ to be aligned with the orbital angular momentum ($0 < \chi_i \leq 1$) or anti-aligned with it ($-1 \leq \chi_i < 0$). By incorporating these dephasing effects into template waveforms, we can potentially detect the presence of new astrophysical objects through observations of a binary's inspiral. As we will show, if the dephasing is large enough, these signals could even be missed by current LIGO/Virgo template-bank searches (see Fig.~\ref{fig:waveform} for an illustration of a dephased waveform).

\vskip 4pt

In principle, the gravitational waves emitted during the binary's merger regime also provide information about the quadrupolar structure of the objects. However, the detailed dynamics of this regime are often sensitive to the microphysics of the objects and can only be resolved accurately through numerical relativity simulations. To preserve analytic control over our waveform (\ref{eqn:waveform}), we will ignore the merger regime throughout this paper. This is achieved in practice by restricting ourselves to low-mass binary systems with $M \lesssim 10 M_\odot$, whose merger frequencies typically lie above the upper bound of the observational windows of ground-based detectors (see \S\ref{sec:peakfreq} later for more detailed discussions). Crucially, these binary systems would inspiral within the detector bands over a large number of orbiting cycles, making matched filtering with our general inspiral waveform the optimal way of searching for them, cf. Fig.~\ref{fig:search_overview}.

\vskip 4pt

While we have only focused on the object's quadrupole moment so far, other types of finite-size effects, such as the object's higher-order spin-induced moments and tidal deformabilities~\cite{Binnington:2009bb, Damour:2009vw, Landry:2015zfa, Pani:2015hfa}, can also contribute to the phase (\ref{eqn:TaylorF2phase}). Nevertheless, these additional terms only start to appear at $3.5$PN~\cite{Levi:2014gsa, Marsat:2014xea, Krishnendu:2017shb} and $5$PN~\cite{nrgr, Flanagan:2007ix} orders, respectively, and are therefore subdominant compared to (\ref{eqn:Kappa2PN}) and (\ref{eqn:Kappa3PN}). In particular, these higher-PN terms are suppressed in the early inspiral of the coalescence, and only become non-negligible in the strong gravity, near merger regime.\footnote{Since the leading-order term in (\ref{eqn:TaylorF2phase}) scales as $\sim v^{-5}$, roughly speaking, orbital parameters that appear at $\lesssim 2.5$PN order affect the phase predominantly in the inspiralling stage, when the number of inspiraling cycles is large, while those with $\gtrsim 2.5$PN become more prominent near merger; see for example Ref.~\cite{Harry:2018hke}.} By focusing on the binary's early inspiral regime we can therefore ignore these higher-order effects. For concreteness, we will set these quantities to their corresponding values for black holes~\cite{Hansen:1974zz, Thorne:1980ru, Landry:2015zfa, Pani:2015hfa}.\footnote{Note however that we retain the $\kappa$-dependence in the 3.5PN phasing term~\cite{Krishnendu:2017shb} in Section~\ref{sec:testbank}.}

\subsection{Cutoff Frequency from Small Compactness} \label{sec:peakfreq}

While the parameter $\kappa$, described in \S\ref{sec:phase}, characterizes the deformation of an astrophysical object's shape, it does not carry information about its size. This is instead described by the compactness parameter
\beq
\mathcal{C} \equiv \frac{m}{r} \, , \label{eqn:compactness}
\eeq
where $r$ is the equatorial radius of the object. Black holes, which are the most compact known astrophysical objects, have $0.5 \leq \mathcal{C} \leq 1$, where the lower and upper bound corresponds to the compactness of a Schwarzschild and an extremal Kerr black hole respectively.
For neutron stars, $\mathcal{C}$ can range between $\sim 0.1 -0.3 $~\cite{Ozel:2016oaf, Abbott:2018exr}. On the other hand, objects that arise in many BSM scenarios can be much more diffuse, with $\mathcal{C} \ll 0.1$ (for example see Refs.~\cite{AmaroSeoane:2010qx, Croon:2018ybs}).

\vskip 4pt

The compactness of each binary component can significantly affect the cutoff frequency of the waveform (\ref{eqn:waveform}) because together they dictate the binary separation at which the merger occurs. For binary systems with highly-compact objects, such as black holes and neutron stars, the merger frequencies must be deduced through detailed numerical relativity studies, as the full non-linear dynamics of the binary merger must be taken into account (see for example Refs.~\cite{Bernuzzi:2015rla, Bohe:2016gbl} for the fitting formulae for the merger frequencies of binary black hole and binary neutron star waveforms). Roughly speaking, these strong-gravity effects become important when the orbital separation, $R$, is smaller than the binary's innermost stable circular orbit (ISCO), $r_{\rm ISCO} \approx 6M$~\cite{Blanchet:2006zz, Schafer:2009dq}. However, if the binary components have sufficiently small $\mathcal{C}$'s, the binary would have already merged at $R < r_{\rm ISCO}$. In this case, the strong-gravity regime is not reached at merger, making the analytic PN approximation still a valid description of the dynamics.

\vskip 4pt

Pledging ignorance to the merger dynamics of binary systems with small-compactness objects, we will terminate our waveform (\ref{eqn:waveform}) when the binary touches, i.e. when $R = r_1 + r_2$.\footnote{Some papers use $R=r_{\rm ISCO}$ as a merger condition for low-compactness binary systems. However, the notion of an ISCO ceases to exist for objects with $\mathcal{C} \lesssim 1/6 \approx 0.17$, as this fictitious ISCO would be located in the interior of the object. This leads to a factor of $\approx 6$ underestimation in $f_{\rm cut}$, which can substantially reduce the frequency range over which the SNR could be accummulated. } Using Kepler's third law, our cutoff frequency is therefore
\beq
f_{\rm cut} = \frac{1}{\pi} \sqrt{\frac{m_1 + m_2}{\left(m_1 / \mathcal{C}_1 + m_2 / \mathcal{C}_2\right)^3}}   \hskip 8pt , \label{eqn:cutoff}
\eeq
where we have ignored PN corrections to (\ref{eqn:cutoff}). Furthermore, we have neglected the influence of the binary components' spins, quadrupoles, and higher-order multipole moments on (\ref{eqn:cutoff}), which are PN suppressed in the early inspiral, but can be important when the binary separation is small.\footnote{For instance, the merger frequencies of compact binary systems in prograde orbits are known to be larger than those in retrograde orbits, see e.g.~\cite{Bohe:2016gbl} for analytic fitting formulae for the relationship between the waveform frequency at peak amplitude and the binary components' spins. However, these effects only appear in the strong-gravity regime and are neglected here for simplicity.} These additional corrections would deepen the gravitational potential of the binary, thereby increasing the cutoff frequency towards larger values~\cite{Blanchet:2006zz, Schafer:2009dq}. In other words, (\ref{eqn:cutoff}) underestimates the actual touching frequency of the binary, and may thus be viewed as a conservative cutoff of our waveform frequency. The impact of this cutoff on our waveform is schematically illustrated in Fig.~\ref{fig:waveform}. In the special case where both of the binary components have the same compactness, $\mathcal{C}_1 = \mathcal{C}_2 = \mathcal{C}$, (\ref{eqn:cutoff}) becomes
\beq
f_{\rm cut} \simeq  1440 \, \text{Hz}  \left( \frac{\mathcal{C}}{0.2} \right)^{3/2} \left(\frac{4 M_\odot}{M}\right)   \, . \label{eqn:cutoff2}
\eeq
For low-mass binary systems ($ M \lesssim 10 M_\odot$), the observational lower bound of ground-based detectors, $f \gtrsim 10 \, \text{Hz} $, implies that we can probe binary inspirals with $\mathcal{C} \gtrsim 10^{-3}$. Although the touching condition (\ref{eqn:cutoff2}) is inaccurate for binaries with $\mathcal{C} \gtrsim 1/6 \approx 0.17$ (see discussion above), their actual merging frequencies are greater than $f \gtrsim 10^3 \, \text{Hz}$, which is beyond the upper bound of the detector sensitivity bands. The precise values of $f_{\rm cut}$ in these cases are therefore immaterial, as the inspiral signal in the observational band remains unchanged. 

\vskip 4pt

Finally, we note that the parameters $\chi, \kappa$ and $\mathcal{C}$ of a given astrophysical object are in principle related to each other. For instance, by requiring the speed of a test mass on the equatorial surface to be smaller than its escape velocity, we can obtain a mass-shedding bound that relates $\mathcal{C}$ with the maximum value of $\chi$. Furthermore, through simple dimensional analysis $Q \propto m r^2$, we find that $\kappa \chi^2 \propto \mathcal{C}^{-2}$. These imply that the dephasing from (\ref{eqn:Kappa2PN}) and (\ref{eqn:Kappa3PN}) are actually correlated with the compactness of the object for a given equation of state. Absent a detailed astrophysical model in mind, we will treat $\chi, \kappa$, and $\mathcal{C}$ as independent parameters, although we emphasize that their implicit correlation can perhaps be exploited in future work, similar to how universal relations~\cite{Yagi:2013bca} are used to simplify analyses of binary neutron star signals.

\section{Detectability of General Inspirals} \label{sec:testbank}

In this section, we study the detectability of our newly proposed waveforms through current matched-filtering searches.
We first construct a binary black hole template bank that is representative of those used by the LIGO/Virgo collaboration (\S\ref{sec:templateBank}). We then investigate how reliable this template bank is at recovering our general inspiral waveforms, specifically by computing its \textit{effectualness} (\S\ref{sec:match}). As a prerequisite to these analyses, we introduce the inner product between two arbitrary waveforms, $h_1$ and $h_2$, defined as~\cite{Cutler:1994ys} 
\beq
  \left(h_1|h_2\right) \equiv 4 \, \mathrm{Re} \int^{\infty}_{0} \d f \, \frac{ \tilde{h}_1(f) \tilde{h}^*_2(f)}{S_n(f)}\, , \label{eqn:innerprod}
\eeq
where $\tilde{h}_1, \tilde{h}_2$ are their Fourier representations, and $S_n$ is the (one-sided) noise spectral density. For future convenience, we define the normalized signal, $\hat{h}_i \equiv h_i/(h_i| h_i)^{1/2}$, such that the normalized inner product is given by
\beq
\left[h_1|h_2\right] \equiv (\hat{h}_1 | \hat{h}_2 ) = \frac{\left(h_1|h_2\right)}{\sqrt{\left(h_1|h_1\right)\left(h_2|h_2\right)}}\, . \label{eqn:dotprodnorm}
\eeq
Throughout this work, we use the \texttt{aLIGO\_MID\_LOW}~\cite{LIGOnoise} detector specification for $S_n$, which is representative of the first LIGO observational run, O1. When evaluating the frequency integral (\ref{eqn:innerprod}), we use the lower and upper cutoff frequencies $f_l=30 \mathrm{\, Hz}$ and $f_u=512 \mathrm{\, Hz}$. These choices reflect the fact that low mass binary inspirals accumulate a minimum of 95\% of their signal-to-noise ratio (SNR) within this frequency range. Finally, although represented as a continuous integral, (\ref{eqn:innerprod}) is in practice evaluated discretely in frequency. We therefore specify a sampling rate of $1024\mathrm{\, Hz}$ and take the maximum time spent in band to be $94 \mathrm{\, s}$.\footnote{This corresponds to the time it takes a binary with component masses $m_1=m_2=1 \, M_{\odot}$, which is the smallest mass we consider in this paper, to inspiral between the stated frequency cutoffs.} These choices give us the grid of possible coalescence times to maximize over when computing the effectualness later.

\subsection{Binary Black Hole Template Bank}
\label{sec:templateBank}

We now construct a template bank that is representative of those used by the LIGO/Virgo Collaboration \cite{Brown:2012qf,DalCanton:2017ala,Roy:2017oul,2018arXiv181205121M}. In order to do so, we use the TaylorF2 waveform model~\cite{Damour:2000zb} for binary black holes~\cite{Arun:2004hn, Wade:2013hoa, Mishra:2016whh} with intrinsic parameters $\bm{p}_{ \tmp} = \{m_1, \, m_2, \,\chi_1, \,\chi_2\}$, where the spins $\chi_1, \chi_2$ are parallel to the orbital angular momentum of the binary. This waveform model is exactly the same as that in (\ref{eqn:waveform}), except we now specify the $\kappa$ parameters to be unity~\cite{Hansen:1974zz, Thorne:1980ru} and neglect the cutoff frequency introduced by the $\mathcal{C}$'s for black holes. Crucially, since we are only interested in the signals emitted during the inspiraling regime, we do not have to use the phenomenological~\cite{Khan:2015jqa} or the effective-one-body waveform~\cite{Bohe:2016gbl} models, which include numerical relativity waveforms near the binary merger.

\vskip 4pt

Matched-filtering searches involve computing the inner product (\ref{eqn:dotprodnorm}) between the data and a set of template waveforms. Practically, this requires a discretized sampling of the parameter space $\bm{p}_{\tmp}$ in the form of a template bank. We adopt the geometric-placement technique described in Ref.~\cite{Roulet:2019hzy}, although many other methods exist; see for example Refs.~\cite{Owen:1995tm, Owen:1998dk, Harry:2009ea,Ajith:2012mn,Brown:2012qf}. 
We refer the reader to the original work for details. The key feature of this formalism lies in the following decomposition of the waveform phase
\beq
 \psi (f; \bm{p}_{\tmp}) = \overline{\psi} (f) + \sum_{\alpha=0}^{N} c_\alpha (\bm{p}_{\tmp}) \, \psi_\alpha (f) \, , \label{eqn:Phase_geometric}
\eeq
where $\overline{\psi} $ is an average behaviour of the phase, which is chosen for convenience, $c_\alpha$ is a set of basis coefficients that only depend on $\bm{p}_{\rm bbh}$, and $\psi_\alpha$ is a set of orthonormal basis functions that satisfy $\langle \psi_\alpha | \psi_\beta \rangle  = \delta_{\alpha \beta}$, with $\langle \cdot | \cdot \rangle$ being an inner product that is slightly modified from (\ref{eqn:innerprod})~\cite{Roulet:2019hzy}. 

\vskip 4pt

To create the basis functions $\psi_\alpha$, we randomly sample $4\times10^4$ parameter combinations and generate waveforms for each. We use the parameter ranges $1.0 M_\odot \leq m_i \leq 3.0 M_\odot$ and $-0.8 \leq \chi_i \leq 0.8$ for component masses and dimensionless spins respectively. This mass range is motivated by the fact that the resulting binary systems have relatively large chirp masses, and at the same time, would inspiral over long periods of time in the detector band (e.g. $\gtrsim 10\,$s). A study which includes a wider mass range, especially subsolar-mass objects, is certainly possible, though it would not alter the qualitative conclusions of this work (see \S\ref{sec:match} for a more detailed discussion). The functions $\psi_\alpha$ are then computed through a singular-value decomposition of a matrix that consists of the phases of the generated waveforms~\cite{Roulet:2019hzy}. We find that taking $N=3$ in (\ref{eqn:Phase_geometric}) is sufficient to describe the behaviour of the phases of these low-mass binary systems (cf. Fig.~\ref{fig:resphase}). The coefficients $c_\alpha(\bm{p}_{\tmp})$ of a given waveform are calculated through the projection $c_\alpha(\bm{p}_{\tmp}) = \langle \psi (\bm{p}_{\tmp}) - \overline{\psi} \, | \, \psi_\alpha\rangle$. For our template bank, we find that the ranges $-953.19 \leq c_0 \leq 326.96$, $-8.18 \leq c_1 \leq 5.76$, and $-0.11 \leq c_2 \leq 0.05$ are sufficient to cover our chosen parameter space. Finally, we take the grid spacing $\Delta c_\alpha = 0.13$\footnote{Not all combinations of $c_\alpha$ give physically realizable waveforms. We therefore use a fudge factor~\cite{Roulet:2019hzy} of $\zeta=0.01$.}, which leads to a total of $562,155$ templates in our bank. As we shall see in \S\ref{sec:match}, these choices lead to a very well-sampled template bank. Note that we did not seek to minimize the number of templates, but instead ensured that our bank's coverage is sufficient to assess the loss of effectualness for our general waveforms.

\subsection{Effectualness to Inspiral Waveforms} 
\label{sec:match}

We are now ready to test how well a binary black hole template bank can be used to detect the general inspiral waveforms that we constructed in Section~\ref{sec:binarywaveform}. For concreteness, we assume that one of the binary components is a black hole, $\kappa_1 = 1$ and $0.5 \leq \mathcal{C}_1 \leq 1$, while the other is a general compact object, whose finite-size parameters are labeled by $\kappa_2 \equiv \kappa$ and $\mathcal{C}_2 \equiv \mathcal{C}$. Since it is important that we distinguish the intrinsic parameters of the binary black hole template waveforms from those of the general waveforms, we will denote them by  $\bm{p}_{\tmp}$ and $\bm{p}_{\gen}$ respectively, with the latter being $\bm{p}_{\gen} = \{ m_1, \, m_2, \, \chi_1, \, \chi_2, \, \kappa, \, \mathcal{C} \}$.

\vskip 4pt

The primary tool for assessing the template bank's effectiveness at recovering our general waveforms is its effectualness~\cite{Damour:1997ub, Buonanno:2009zt}. More precisely, this is obtained by maximizing the inner product (\ref{eqn:dotprodnorm}) between the template and general waveforms over their relative time of coalescence $t_c$, phase of coalescence $\phi_c$, and the intrinsic parameters $\bm{p}_{\tmp}$ of every template in the bank:
\beq
\varepsilon \left( \bm{p}_{\tmp}, \bm{p}_{\gen} \right) \equiv \max _{t_{c}, \phi_{c}, \{\bm{p}_{\tmp} \} } \hskip 2pt \left[h (\bm{p}_{\tmp}) \, | \, h (\bm{p}_{\gen}) \right]  \, , \label{eqn:effectualness}
\eeq
where $\{ \bm{p}_{\tmp} \}$ denotes the list of template parameter combinations. In other words, (\ref{eqn:effectualness}) quantifies the overlap between the general waveform and the best-fitting template waveform in the bank. To obtain a rough idea of how a reduced effectualness translates into a loss in signal detectability, we note that existing searches adopt an SNR detection threshold of 8, while a typical binary system detected thus far has an SNR ranging from $10$ to $15$~\cite{LIGOScientific:2018mvr}. For a signal with true SNR of $12.5$, a template bank with effectualness $ \varepsilon < 8/12.5 \approx 0.64$ would reduce the observed SNR to values below the detection threshold, thereby leading to missed events. This can instead be phrased as a reduced sensitive volume of $1-\varepsilon^3 \gtrsim 0.74$~\cite{Lindblom:2008cm}. A commonly adopted target in LIGO and Virgo searches is $\varepsilon > 0.97$, which leads to a $10\%$ loss in sensitive volume.\footnote{Note that the matched filter SNR alone is not enough to quantify the significance of an event for the search pipelines used by the LIGO and Virgo collaborations. Instead, the false alarm rate and probability of an event being of astrophysical origin ($p_{\rm astro}$) are also used~\cite{LIGOScientific:2019hgc}. Unlike the SNR, these metrics account for the non-stationary and non-Gaussian behaviours of the detector noise and measure the consistency of the potential signal with the best fit waveform. Nevertheless, for the optimal case of stationary Gaussian noise, which we assume in this work by adopting the \texttt{aLIGO\_MID\_LOW}~\cite{LIGOnoise} detector noise curve, the SNR is sufficient to quantify the degraded sensitivity of searches for general inspirals.} 

\vskip 4pt

While the maximization of (\ref{eqn:effectualness}) over $t_c$
and $\phi_c$ can be performed efficiently~\cite{Schutz:1989cu}, the iterative computation over the list $\{  \bm{p}_{\tmp} \}$ is much more computationally expensive. One of the benefits of the geometric-placement method described in \S\ref{sec:templateBank} is that this brute-force iteration can be replaced by a simple search for the \textit{best-fitting point, $ \left\{ \hskip 1pt c_\alpha (\bm{p}_{\tmp}) \hskip 1pt \right\}_{\mathrm{best}}$}, in the bank~\cite{Roulet:2019hzy}. This is achieved by projecting $c_\alpha(\bm{p}_{\gen}) = \langle \psi (\bm{p}_{\gen}) - \overline{\psi} \, | \, \psi_\alpha\rangle$, where $\psi (\bm{p}_{\gen})$ is the phase of the general waveform, while $\overline{\psi}$ and $\psi_\alpha$ are the average phase and basis functions constructed for our bank in (\ref{eqn:Phase_geometric}), respectively. The best-fitting point  $ \left\{ \hskip 1pt c_\alpha (\bm{p}_{\tmp}) \hskip 1pt \right\}_{\mathrm{best}}$ is the closest $c_\alpha(\bm{p}_{\tmp})$ to $c_\alpha(\bm{p}_{\gen})$, as measured by their Euclidean distance. The effectualness can then be evaluated straight-forwardly using these nearby parameters and maximizing over $t_c$ and $\phi_c$.

\vskip 4pt

\begin{figure}[t!]
        \centering
        \includegraphics[width=\textwidth]{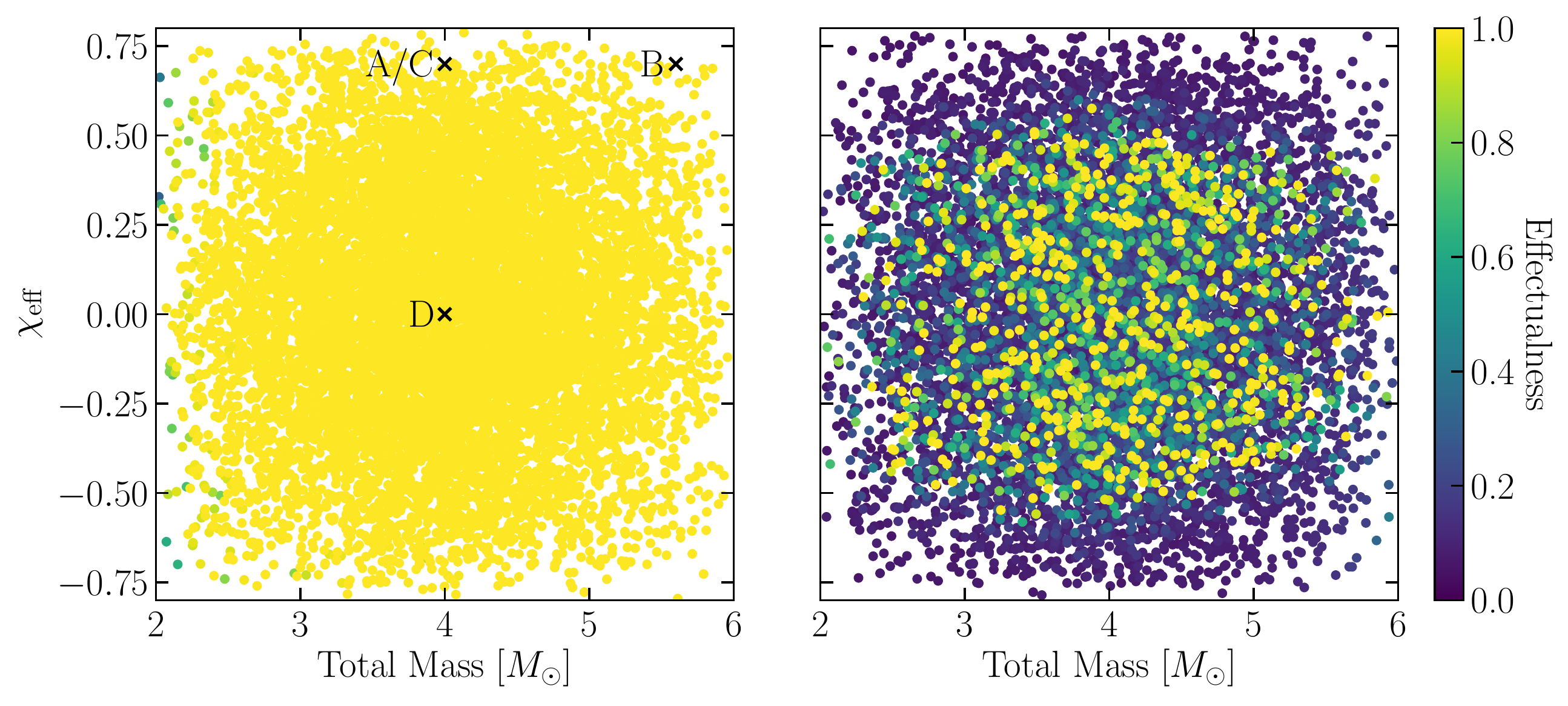}
\caption{
Effectualness of our template bank for binary black hole inspiral waveforms (\textit{left}) and general inspiral waveforms with $\kappa=500$ and $\mathcal{C}=0.1$ (\textit{right}). The effectualness is defined in (\ref{eqn:effectualness}). Comparing these panels we see a drastic loss in effectualness from the finite-size effects. For convenience, we have indicated the masses and spins of scenarios A$-$D that are considered in Table~\ref{table:scenarios} and Fig.~\ref{fig:matchvkap} in the \textit{left} panel. Note that all points are plotted from least to most effectual --- the points with the highest effectualness are therefore the most visible.
\label{fig:tbankeffectualness} } 
\end{figure}

To test the validity of this procedure, we randomly sample an independent set of $10^4$ binary black hole waveforms within the same parameter ranges used to generate $\psi_\alpha$ in \S\ref{sec:templateBank}. We then compute our template bank's effectualness to these waveforms with the prescription above. The result is shown in the \textit{left} panel of Fig.~\ref{fig:tbankeffectualness}, where the effectualness is plotted as a function of the binary total mass $M$ and the effective mass-weighted spin, $\chi_{\rm eff} = \left( m_1 \chi_1 + m_2 \chi_2 \right) / M $.  We find that 99\% of the random templates have $\varepsilon > 0.97$.
We also find that the effectualness decreases slightly as $\chi_{\rm eff}$ increases, indicating that the basis functions are less able to capture the high-spin behaviour. In essence, this figure demonstrates both the validity of our method of evaluating (\ref{eqn:effectualness}) and our construction of a highly effectual bank for detecting binary black hole signals.

\vskip 2pt

\begin{figure}[t!]
        \centering
        \includegraphics[scale=0.7]{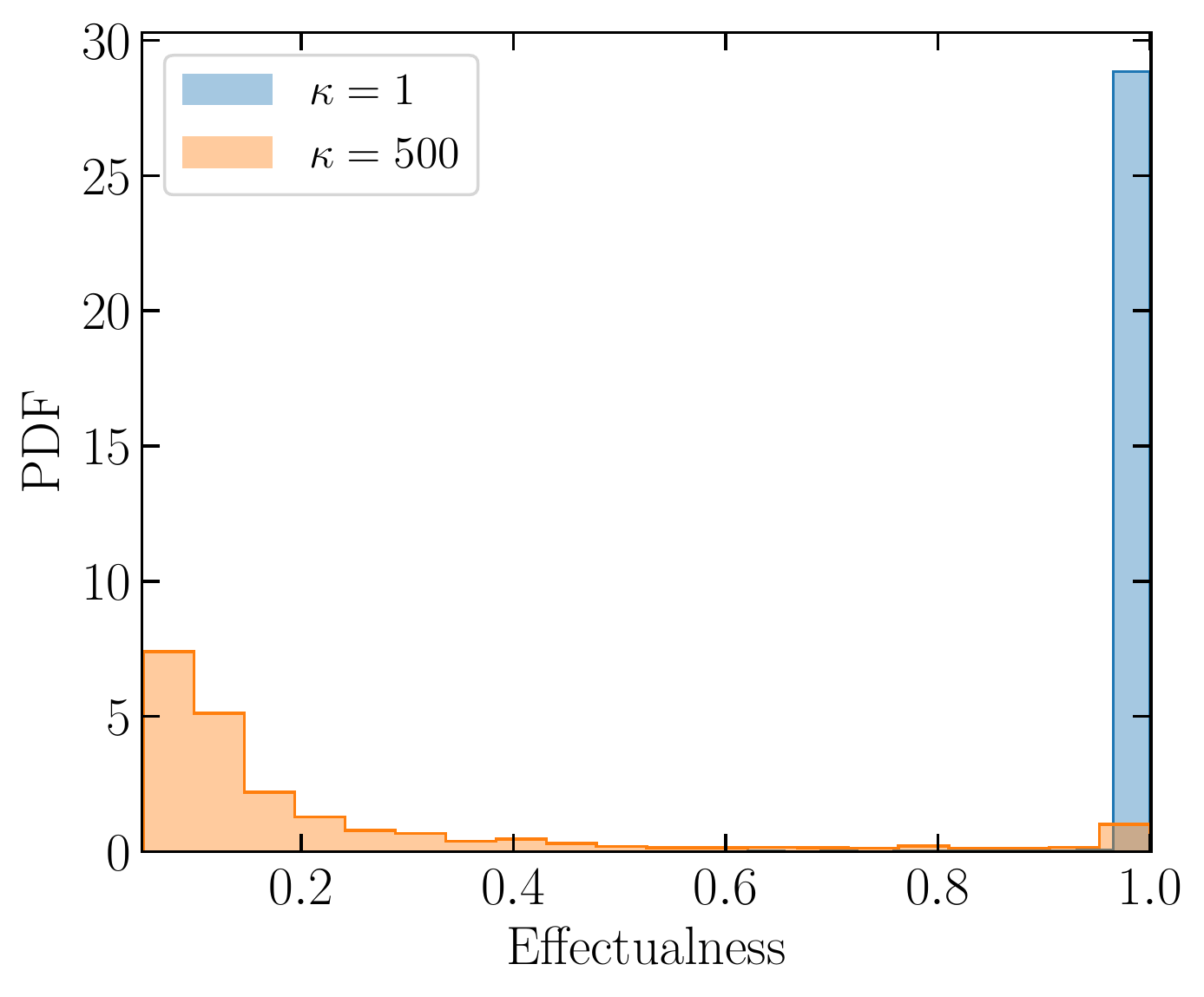}
\caption{Probability distribution functions (PDFs) of the randomly generated $\kappa=1$ and $\kappa=500$ waveforms shown in Fig.~\ref{fig:tbankeffectualness}.\label{fig:PDF} } 
\end{figure}

Taking the \textit{left} panel of Fig.~\ref{fig:tbankeffectualness} as a baseline (optimal) effectualness of our template bank, we can compare the bank's effectualness to a general inspiral waveform. For concreteness, we generate $10^4$ general inspiral waveforms within the same mass and spin ranges, fixing $\kappa=500$ and $\mathcal{C} = 0.1$. The effectualness is shown in the \textit{right} panel of Fig.~\ref{fig:tbankeffectualness}, where we see that a large spin-induced quadrupole moment can significantly decrease the effectualness of the bank. This is especially true in the large-spin limit, since the phase contributions (\ref{eqn:Kappa2PN}) and (\ref{eqn:Kappa3PN}) are proportional to $ \kappa_i \chi_i^2$.  Statistically, we find that only 6\% of the random signals have $\varepsilon > 0.9$ and 29\% have $\varepsilon > 0.2$.  This degradation in effectualness is entirely due to the spin-induced quadrupole, as the cutoff frequency of the waveform for $\mathcal{C}=0.1$ is greater than $f_u$, thereby having no effect on our analysis. The probability distribution functions (PDFs) for the waveforms in Fig.~\ref{fig:tbankeffectualness} are shown in Fig.~\ref{fig:PDF}.

\begin{table}[h!]
\centering
\begin{tabular}{|c|ccccc| l|}
\hline
Scenario & $m_{1} \,[\mathrm{M_{\odot}}]$ & $m_2\,[\mathrm{M_{\odot}}]$ & $\chi_1$ & $\chi_2$ & $\mathcal{C}$ & Description\\
\hline
A & 2.0 & 2.0 & 0.7 & 0.7 & 0.1 &  Fiducial case \\
B & 2.8 & 2.8 & 0.7 & 0.7 & 0.1 &  Heavier total mass \\
C & 3.0 & 1.0 & 0.7 & 0.7 & 0.1 &  Lighter general object \\
D & 2.0 & 2.0 & 0.2 & -0.2 & 0.1 &  Reduced and anti-aligned spins \\
E & 2.0 & 2.0 & 0 & 0 & 0.01 &  Reduced compactness \\
\hline
\end{tabular}
\caption{List of representative binary configurations for Fig.~\ref{fig:matchvkap}. The parameters $\{ m_1, \chi_1 \}$ describe the black hole, while $\{ m_2, \chi_2, \kappa, \mathcal{C}\}$ are the parameters of the general astrophysical object.}
\label{table:scenarios}
\end{table}%

\newpage

\begin{figure}[t!]
        \centering
        \includegraphics[scale=0.7]{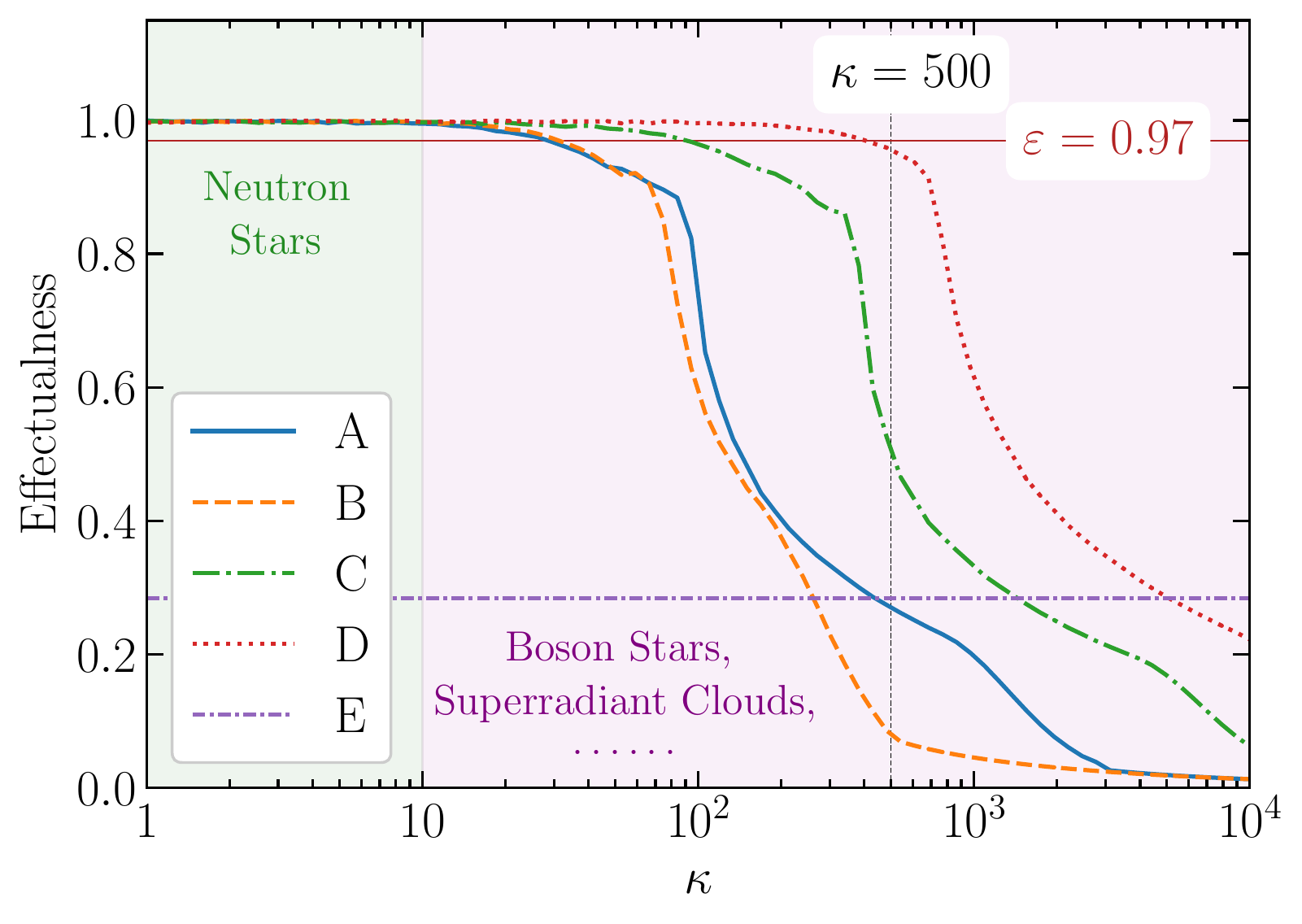}
\caption{Effectualness of the various scenarios listed in Table~\ref{table:scenarios} as a function of $\kappa$. The vertical dashed line denotes systems with $\kappa=500$ and is included for comparison with the \textit{right} panel of Fig.~\ref{fig:tbankeffectualness}. We also include the reference line $\varepsilon=0.97$, which is a commonly adopted requirement for template banks in actual searches. For convenience, we indicate the approximate ranges of $\kappa$ for neutron stars and various BSM objects in green and purple respectively. Kerr black holes have $\kappa=1$.}
    \label{fig:matchvkap}
\end{figure}

Instead of fixing $\kappa$ and $\mathcal{C}$, it is also interesting to examine the bank's effectualness as a function of these parameters. For concreteness, we consider five qualitatively-distinct scenarios, which are listed in Table~\ref{table:scenarios}. For each case, we treat $\kappa$ as a free parameter and calculate the effectualness within the range $1 \leq \kappa \leq 10^4$. Note that scenario E has no spins and is therefore unaffected by varying $\kappa$; this scenario is meant to show the effect of reducing the compactness of an object, which terminates the waveform in the detector sensitivity band. The results are presented in Fig.~\ref{fig:matchvkap}, where we find that each scenario with non-vanishing spins (A$-$D) shows a general behaviour with the following three distinct regions as a function of $\kappa$:
\begin{enumerate}
    \item No loss in effectualness at low values of $\kappa$. For scenario A this occurs for $1 \leq \kappa \lesssim 20$.
    
    \item We see a series of different rates of declining effectualness as a function of $\kappa$, possibly indicating the varying rates of importance of the various higher-order PN contributions such as (\ref{eqn:Kappa2PN}) and (\ref{eqn:Kappa3PN}). For scenario A we see this at $20\lesssim \kappa \lesssim 3\times10^3$.
    
    \item Finally we see a flattening of the effectualness. This flattening occurs when $\kappa$ is so large that the 2PN term (\ref{eqn:Kappa2PN}) becomes the dominant contribution to the phase evolution. At this point, maximising over $t_c$ and $\phi_c$ always finds a small region of frequency space where the overlap between the two waveforms is nearly vanishing.
\end{enumerate}
We tested many additional scenarios and found this overall behaviour to be universal, showing the three distinct regions described above. Note that the value of $\kappa$ at which each scenario enters the three regions and the length spent there differs greatly as can be seen in Fig.~\ref{fig:matchvkap}. Below we give some qualitative arguments to compare the differences between the scenarios A$-$E.

\vskip 4pt

Importantly, in the first region we see no noticeable reduction in the effectualness for $ \kappa \lesssim 20$, even for reasonably highly-spinning objects. As mentioned in \S\ref{sec:phase}, this range overlaps with the expected values of $\kappa$ for neutron stars~\cite{Laarakkers:1997hb, Pappas:2012ns}. It is for this reason that $\kappa$ can be safely ignored when searching for black hole - neutron star systems with binary black hole templates, although variations from $\kappa=1$ must be accounted for during parameter estimation \cite{Abbott:2018exr}. Binary neutron star systems would have additional contributions to their phase evolution since both objects now contribute to the phase with $\kappa \gtrsim 1$. LIGO and Virgo typically only consider slowly spinning neutron stars, $\chi \leq 0.4$ \cite{Ajith:2012mn}, where the effect of $1 < \kappa \lesssim 10$ is small --- binary black hole templates are therefore still suitable at the search level.

\vskip 4pt

For larger values of $\kappa$, the effectualness quickly drops below the normal requirement of $\varepsilon \geq 0.97$ for template banks. Scenarios A and B show similar behaviour up to $\kappa \approx 200$ at which point they start to diverge. To understand this behaviour, we first note that the prefactors of the $v-$dependence in (\ref{eqn:Kappa2PN}) and (\ref{eqn:Kappa3PN}) are unchanged for equal-mass-ratio binary systems, regardless of their total mass. However, $v$ retains some dependence on the total mass --- $v\propto M^{1/3}$. The various PN terms therefore scale differently with $M$, causing a different rate of loss of effectualness. The difference between scenarios A and C can again be easily understood by examining the prefactors of (\ref{eqn:Kappa2PN}) and (\ref{eqn:Kappa3PN}). Since our general object is the lighter of the two components in scenario C (see Table~\ref{table:scenarios}), the mass dependencies of these terms dictate that the general object contributes less to the phase evolution. This reduced contribution can be compensated for by a larger value of $\kappa$, producing an overall shift to the right in Fig.~\ref{fig:matchvkap} from scenarios A to C. Similarly, if we were to fix the total mass and choose the heavier component to be our general object, we would see an overall shift from scenario A to the left. Finally, scenario D has significantly smaller spins, reducing the overall phase contribution from the spin-induced quadrupole.

\vskip 4pt

For scenario E we see a significant reduction in effectualness, even for small values of $\kappa$ (the horizontal line merely reflects our choice of vanishing spins). This is because the frequency cutoff set by (\ref{eqn:cutoff}) is $f_{\rm cut} \approx 44 \, \mathrm{Hz}$, which lies inside the sensitivity band of ground-based detectors. For comparison, the cutoffs for scenarios A$-$D lie above $f_u=512 \, \mathrm{Hz}$ in our analysis. More generally, we find that  $\mathcal{C} \gtrsim 0.05$ gives a cutoff frequency of $f_{\rm cut} \gtrsim f_u$.\footnote{ The precise range of $\mathcal{C}$ depends on the mass of the binary components, and can be calculated more accurately through (\ref{eqn:cutoff}).} This loss in effectualness cannot be compensated for by adding additional waveforms to the bank, unlike for scenarios A$-$D. Instead it represents a truncation of the waveform and therefore a reduction in SNR. Since the introduction of $f_{\rm cut}$ merely represents our ignorance of the actual merger dynamics of the binary, the effectualness for objects with small values of $\mathcal{C}$ can potentially be improved. For instance, as discussed in \S\ref{sec:peakfreq}, PN corrections to (\ref{eqn:cutoff}) would increase $f_{\rm cut}$ towards higher values. Alternatively, model-dependent numerical relativity simulations can be performed to fully extract the merger waveforms. 


It is important to assess whether the loss of effectualness in Fig.~\ref{fig:matchvkap} is due to the limited range of component masses considered in the bank. We study this by repeating the procedure outlined in \S\ref{sec:templateBank} but instead consider a component mass range $1.0 M_\odot \leq m_i \leq 5.0 M_\odot$. We find no difference in the initial reductions of effectualness for all scenarios (this corresponds to the $\kappa \lesssim 100$ and $\varepsilon \gtrsim 0.85$ region in scenario A). As $\kappa$ increases to larger values, the effectualness still drops rapidly (as observed for $\kappa \gtrsim 100$ in scenario A) although the rates of reduction slightly decrease. This slight improvement occurs because, in this large-$\kappa$ region, the best-fitting points $ \left\{ \hskip 1pt c_\alpha (\bm{p}_{\tmp}) \hskip 1pt \right\}_{\mathrm{best}}$ are located inside and outside the bounds of $c_\alpha(\bm{p}_{\tmp})$ in the bigger and smaller template bank, respectively. In contrast, the earlier reduction is robust to the increased component mass range because $ \left\{ \hskip 1pt c_\alpha (\bm{p}_{\tmp}) \hskip 1pt \right\}_{\mathrm{best}}$ lies within the bounds of $c_\alpha(\bm{p}_{\tmp})$ for both banks. Despite this slight dependence on the size of the template bank parameter space, the effectualness maintains its monotonically-decreasing trend with increasing $\kappa$ for all scenarios. This robust behaviour suggests that waveforms with large $\kappa$ cannot be mimicked by binary black hole waveforms with vastly wrong intrinsic parameters.

\begin{figure}[t!]
        \centering
        \includegraphics[width=\textwidth, trim= 20 0 0 0]{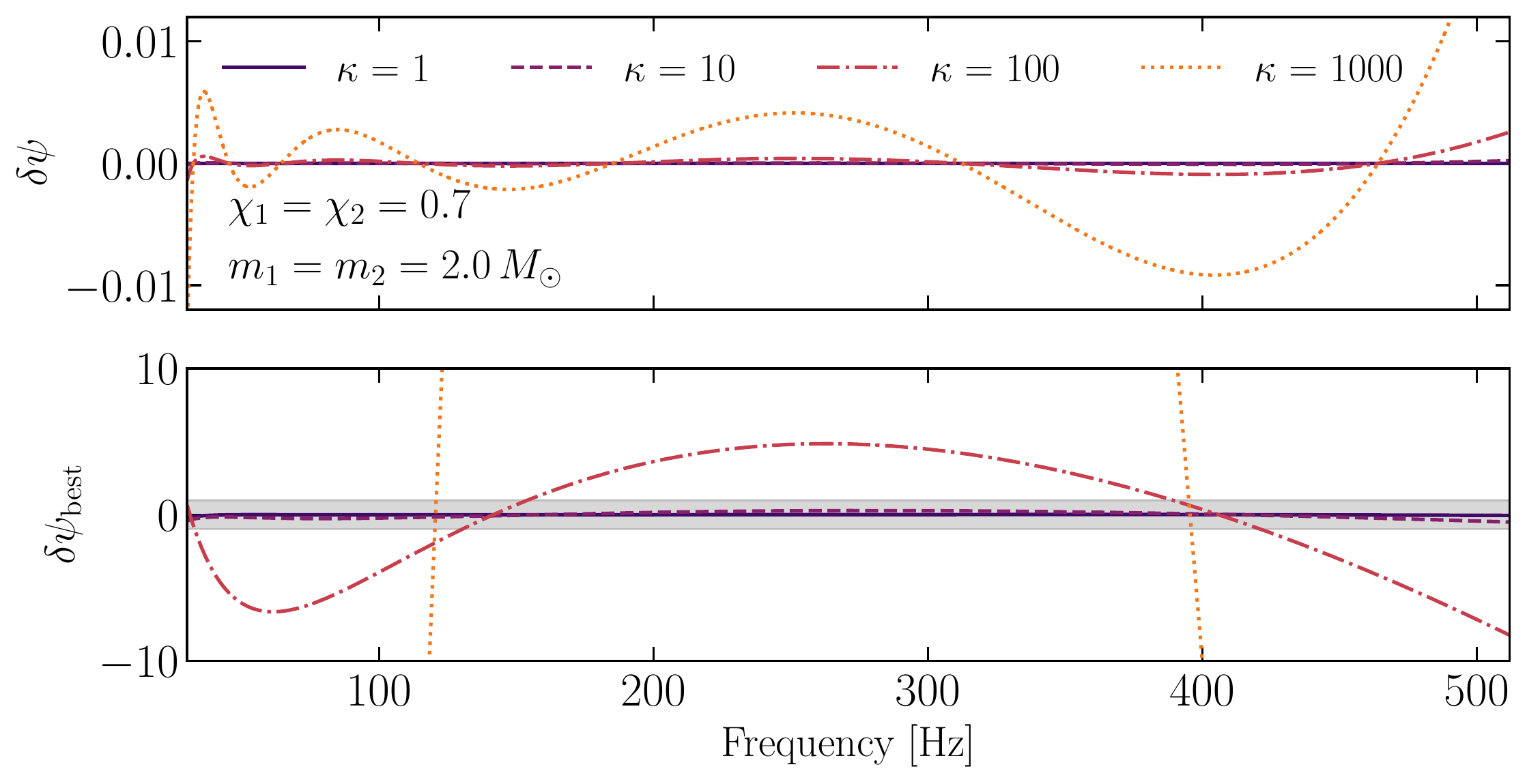}
\caption{The residual phases of scenario A at various values of $\kappa$. In the \textit{top} panel, the magnitudes of $\delta \psi$ are smaller than unity, indicating that the basis functions $\psi_\alpha$ can sufficiently describe the phases of the general waveforms. The gray band in the \textit{bottom} panel indicates the range  $ -1 \leq \delta \psi_{\mathrm{best}} \leq 1$. The phase residual $\delta \psi_{\mathrm{best}}$ for $\kappa \gtrsim 100$ clearly exceeds order unity, resulting in a reduced effectualness of the binary black hole template bank.}
    \label{fig:resphase}
\end{figure}

\vskip 4pt
 
While the effectualness is a good measure of the differences between the signal and template waveforms, it is an integrated quantity with a less clear interpretation. We therefore examine two phase residuals:
\beq
\begin{aligned}
 \delta \psi  = \psi(\bm{p}_{\gen}) - \left[\overline{\psi} + \sum_{\alpha=0}^{N=3} c_\alpha (\bm{p}_{\gen}) \, \psi_\alpha \right]\, , 
 \quad 
 \delta \psi_{\mathrm{best}}  = \psi(\bm{p}_{\gen}) - \left[\overline{\psi} + \sum_{\alpha=0}^{N=3} \left\{ \hskip 1pt c_\alpha (\bm{p}_{\tmp}) \hskip 1pt \right\}_{\mathrm{best}} \, \psi_\alpha \right]\, ,\label{eqn:Phase_residual}
 \end{aligned}
\eeq
where $c_\alpha(\bm{p}_{\gen}) = \langle \psi (\bm{p}_{\gen}) - \overline{\psi} \, | \, \psi_\alpha\rangle$ (see \S\ref{sec:templateBank}). The residual phase $\delta \psi$ is a measure of the basis functions $\psi_\alpha$'s ability to capture the phase evolution of a general waveform. Instead, $\delta \psi_{\mathrm{best}}$ quantifies the phase deviation between the general waveform and its best-fitting waveform in our bank.

\vskip 4pt

These residual phases are shown in Fig.~\ref{fig:resphase} at various values of $\kappa$ for scenario A. In the \textit{top} panel, the fact that $\left| \delta \psi \right| \ll 1$ for all values of $\kappa$ indicates that our basis functions, with $N=3$ in (\ref{eqn:Phase_geometric}), are able to describe the phase of the general waveform to a high degree of accuracy. The reduction in effectualness in Fig.~\ref{fig:matchvkap} is therefore a consequence of the large separations between $c_\alpha(\bm{p}_{\gen})$ and $\left\{ \hskip 1pt c_\alpha (\bm{p}_{\tmp}) \hskip 1pt \right\}_{\mathrm{best}}$. In the \textit{bottom} panel, we see that $\left| \delta \psi_{\mathrm{best}} \right| \ll 1$ for $\kappa=1 $ and $\kappa=10 $ --- it is for this reason that we can still detect these binary systems with standard binary black hole waveforms, in agreement with Fig.~\ref{fig:matchvkap}.  For $\kappa =100$, we start to see deviations exceeding order unity, $\left| \delta \psi_{\mathrm{best}} \right| \gtrsim \mathcal{O}(1)$, resulting in a reduced effectualness. Crucially, the \textit{bottom} panel of Fig.~\ref{fig:resphase} illustrates that our general inspiral waveforms are \textit{not degenerate} with a binary black hole template waveform with the wrong intrinsic parameters; if this were the case, we would see zero phase residuals.

\vskip 4pt

In a nutshell, Figs.~\ref{fig:tbankeffectualness},~\ref{fig:matchvkap}, and \ref{fig:resphase} clearly indicate that, for values of $\kappa \lesssim 20$, binary black hole template banks are still able to detect general astrophysical objects. On the other hand, for $\kappa \gtrsim 20$, there are large parts of parameter space where the binary black hole templates cannot be used to recover these general signals. This is especially true if the general object is highly spinning and has a larger mass when compared with its binary companion. Moving forward, new template banks must be constructed with $\kappa$ as a one-parameter extension to the standard waveforms. Since $\mathcal{C}$ is simply a truncation of the waveform, it is not necessary to include it as an additional parameter.

\section{Summary and Outlook} \label{sec:conclusion}

In this paper, we examined whether binary black hole template banks can be used to search for the gravitational waves emitted by a general binary coalescence. We focused on binary systems with components that can have large spin-induced quadrupole moments and/or small compactness. Figure~\ref{fig:matchvkap} clearly demonstrates that as the quadrupole term becomes large, its phase contribution to the waveform becomes significant. Binary black hole template banks are thus insufficient for searching for these general astrophysical objects. More precisely, we find that the effectualness of these template banks are quickly reduced for $\kappa \gtrsim 20$ for highly-spinning objects (for example scenario A in Table~\ref{table:scenarios}). This range of $\kappa$ coincides with an interesting part of the parameter space where compact objects in various BSM scenarios may exist~\cite{Baumann:2018vus, Baumann:2019ztm, Ryan:1996nk, Herdeiro:2014goa}. Figure~\ref{fig:resphase} further shows that these signatures are not degenerate with binary black hole template waveforms with the wrong intrinsic parameters. It is therefore essential that extended template banks are created in order to search for these novel signatures. As a byproduct of our analysis, we recovered the result that the effectualness remains high for smaller values of $\kappa$. Binary black hole waveforms can therefore be used to search for binary systems with neutron stars, as is currently done by the LIGO/Virgo collaboration~\cite{TheLIGOScientific:2017qsa}.


In addition, we considered the impact of an object's compactness on the merger frequency of the binary. Since a detailed description of the merger dynamics is model-dependent, we truncate the waveform through a frequency cutoff. For objects with small-compactness, this cutoff is set by the point at which the binary components touch. This truncation only has a significant effect on the effectualness when the cutoff frequency is within the sensitivity bands of ground-based detectors. As a fiducial guide, our estimate shows that this is the case for $\mathcal{C} \lesssim 0.05$ in low-mass binary systems. This loss in effectualness can be compensated for through more detailed modeling of the binary merger dynamics.

\vskip 4pt

Throughout this paper, we focused on the early inspiral regime of low-mass binary systems. This restriction had multiple benefits. Firstly, the inspiral is a regime where analytic results of the PN dynamics are readily available. This provided us with a well-defined framework to construct our general waveforms, where the physics contributing to waveform deformations can be clearly interpreted. Secondly, ground-based detectors are able to probe these inspirals over hundreds or thousands of cycles, thereby allowing for a precise characterization of the physics at play. Inspiral signals in LIGO/Virgo observations therefore represent a new avenue to probe BSM physics and novel astrophysical phenomena.

\vskip 4pt

Our findings show that many new signatures could be missed by current search pipelines.  Although we focused on finite-size effects, many other types of physical phenomena can affect the frequency evolution of a binary. We hope to incorporate these additional dynamics into our general waveforms in future work. Furthermore, we aim to search for these novel signatures in the data collected in the O1$-$O3 observation runs. Using the same procedure and grid spacing as in \S\ref{sec:templateBank}, we estimate that an order-of-magnitude more templates would be required to search for these new signals, though we leave a more refined study to future work. While a detection in these data would certainly indicate a signature of new physics, non-observations can also be used to place meaningful bounds on the space of astrophysical objects that exist in our Universe. Importantly, we believe that this paper represents the first concrete step towards our goal of utilizing gravitational wave inspirals to search for novel astrophysical phenomena and physics beyond the Standard Model.

\subsection*{Acknowledgements}

We thank Liang Dai, Tanja Hinderer, Cody Messick, Samaya Nissanke, Rafael Porto, John Stout, Tejaswi Venumadhav,  Christoph Weniger, Matias Zaldarriaga and Aaron Zimmerman for helpful discussions.  The work of HSC is supported by the Netherlands Organisation for Scientific Research (NWO). TE acknowledges support by the Vetenskapsr\r{a}det (Swedish Research Council) through contract No.  638-2013-8993 and NWO through the VIDI research program ``Probing the Genesis of Dark Matter" (680-47-532; TE). HSC thanks the Oskar Klein Centre at Stockholm University for its hospitality while some of this work was completed. TE thanks the Weinberg Theory Group at the University of Texas at Austin for its hospitality while some of this work was completed. We also thank the authors of Ref.~\cite{Roulet:2019hzy} for sharing their template bank generation code. Numerical computations and plots in this paper are produced with the Python scientific computing packages NumPy~\cite{numpy} and SciPy~\cite{scipy}. This research utilised the HPC facility supported by the Technical Division at the Department of Physics, Stockholm University.


\newpage
\phantomsection
\addcontentsline{toc}{section}{References}
\bibliographystyle{utphys}
\bibliography{main}
\end{document}